\documentstyle[11pt,titlepage]{article}

\begin{document}
\baselineskip 17pt
\newcommand{\be}{\begin{equation}}
\newcommand{\en}{\end{equation}}
\def\ltsima{$\; \buildrel < \over \sim \;$}
\def\lsim{\lower.5ex\hbox{\ltsima}}
\def\gtsima{$\; \buildrel > \over \sim \;$}
\def\gsim{\lower.5ex\hbox{\gtsima}}
\newcommand{\ggg}{$\gamma$}
\newcommand{\eee}{$e^{\pm}$}
\newcommand{\lap}{$L_{38}^{-1/3}$}
\newcommand{\ergs}{\rm \ erg \; s^{-1}}
\newcommand{\msol}{\su M_{\odot} }
\newcommand{\etal}{et al.\ }
\def\msole {~M_{\odot}}
\def\rref{\par\noindent\hangindent=1.5truecm}

\vspace*{1.in}
\noindent

\vspace*{2.in}

\centerline{\bf \Large The WAXS/WFXT MISSION}

\vskip .3in
\centerline{\large Guido Chincarini}

\vskip .3in
\centerline{Osservatorio Astronomico di Brera \& Universit\`a degli Studi di Milano}

\vskip .3in
\centerline{On behalf of the WAXS/WFXT team whose main contributors are listed in 
the acknowledgements}
  
\vskip .3in

\centerline{\bf \large Abstract}

\bigskip

I present the science goals and give a brief summary of the Wide Angle X-ray survey with a Wide 
Field X-ray Telescope (WAXS/WFXT) mission proposal (Phase A) which will be submitted to 
the Italian Space Agency (ASI) following the call for proposal under the Small Satellite program. 
The text points out the uniqueness of the mission for the study of the evolution of clusters of 
galaxies and of the Large-Scale Structure at large redshifts and for the study of the Milky Way. I 
present, furthermore, the successful result of the metrology of the first wide field X-ray optics 
ever made.

\newpage

\tableofcontents

\vskip .2in
\section{Introduction}

The idea of planning a survey aimed especially to the detection of clusters of galaxies came long 
ago following a suggestion by Riccardo Giacconi. At that time Riccardo, in collaboration with 
Chris Burrows and Richard Burg, was studying the possibility to design an X-Ray optics having a 
good resolution over a large field of view. This would optimize the time it would take to carry out 
a survey over a large solid angle. Once it had been demonstrated that the design was under 
control, we started to develop the technology to realize the prototype. This took some years and 
finally we succeeded. 
In these proceedings we will briefly outline the science goals and the general plan of the mission. 
Further details are given in the study of phase A that will be submitted to the Italian Space 
Agency in November 1998 and on specialized papers published especially by the technology X-Ray 
group headed by Oberto Citterio. At this meeting, however, we present for the first time the 
excellent results of the X-Ray metrology carried out on the 60-cm shell prototype, the most 
difficult and critical shell to be made. This is a breakthrough in the field comparable, in all 
aspects, to the result of the first Ritchey Chretien optics for ground based optical telescopes.

WAXS/WFXT is an excellent and unique survey mission with a strong Italian heritage. The ROSAT 
all sky survey is too shallow and the ROSAT deep surveys have too small a solid angle. AXAF will not 
be dedicated to large surveys and does not have the field of view to discover a sizeable number of 
objects. The disadvantage of the XMM serendipitous survey is that it will be spread over thousands of 
pointing in different directions, this is not suitable for measuring the Large Scale Structure. ABRIXAS, 
while being a complement to the proposed mission, will not have the adequate sensitivity and angular 
resolution for our science goals. (The rather limited resolution makes identification directly from the 
survey difficult and places a heavy demand on the telescope time needed for the optical follow-up). 
WAXS will complement the above missions by accomplishing original science and by creating a 
unique catalogue for follow-up observations. Both large ground-based telescopes and space missions 
will make use of the WAXS source catalogues for many years to come.

A comparison with the most important missions that are ready to fly is significant and illuminating. 
This is shown in Fig. I.1, where we plot the area of the survey versus the limiting sensitivity. The 
baseline of the mission includes an ultra-deep survey that almost reaches the sensitivity of the AXAF 
deep survey, but over an area more than twenty times larger. The confusion limit is, assuming we reach 
the optimum resolution as we expect, well below the XMM confusion limit.
 
The work I am describing below is the result of the creativity and dedication of many scientists, 
Italian and Foreign Institutions. To them goes the merit of the content  and however I am 
responsible for the form and the eventual inaccuracies of the text. The main contributors will be 
acknowledged at the end of the contribution. I would like from the starting, however, to express 
my gratitude to the team of the Brera Observatory whom, especially in this last year, worked with 
extreme dedication to this project. The collaboration of Steeve Murray, Alan Wells and Cosimo 
Chiarelli went beyond duty and could be explained only as a result of true friendship and very 
deep interest in the project. 

I will discuss part of the science goals in section II and I will illustrate the expected performance 
of the instrument in section III. In section IV I will give a brief summary of the mission planning. 
At the time of writing it is not yet known whether the mission will be approved. I hope to 
convince the reader in that {\bf this mission is a unique opportunity to extend our understanding 
of the Structure and Origins of the Cosmos}.

\vskip .2in
\section{Science goals}

The unique features of the X-ray sky make it possible to select groups, clusters, and AGN from X-ray 
images and to use these classes of objects to map the large scale structure of the Universe at high 
redshift ($z>1$). Compared to optical images, X-ray images are relatively sparse and dominated by the 
distant, extragalactic sources.

A convincing example is given in Fig. II.1, where we 
reproduce a patch of sky from the second generation Digitised Sky Survey plates, $30'\times30'$ in 
size and corresponding to a ROSAT-PSPC pointed observation (targeted at QSO1404+286).  In 
this optical image there are 2176 objects brighter than the plate limit $m_B\sim23$, about half 
of which are galaxies.   Contours from the ROSAT X-ray image are overlaid on the optical 
picture.  The X-ray data reveals two clusters, `A' at $z=0.36$ and `C' at $z=0.55$, and a group 
of galaxies, `B', at $z=0.12$.  About 20 of the 26 detected sources are distant AGNs. 

{\bf The design 
of the WAXS/WFXT mission is indeed based on the scientific goal of detecting clusters of 
galaxies at high redshifts}. Optical surveys are extremely limited, even if complementary, 
in detecting clusters. This is because for distant clusters optical surveys generally detect 
only the very tip of the luminosity function, that is the brightest galaxies. The galaxy 
background, however, increases tremendously with distance so that the cluster galaxies are 
confused in the background and clusters are difficult to detect even at $z\sim0.8-1.0$.  In the 
X-ray, clusters of galaxies appear as extended objects and an angular resolution of 15 
arcseconds over a large field of view is sufficient to separate clusters from point sources 
at large redshift. In any cosmology the minimum angular diameter of a cluster is about 30 
arcseconds, occurring at about $z\sim1.25$. 

Finally it is important to detect clusters at various redshifts and over a very large area. 
This is necessary to obtain a statistically significant number of bright clusters, which 
are very rare, and to study their evolution and clustering properties. Thus, the primary 
mission requirement is to conduct two surveys. A 900 square degree shallow survey to detect 
the brightest clusters over a large solid angle of contiguous area, and a 100 square degree
deep survey to probe more deeply into the cluster X-ray luminosity function to study evolution. 

In Fig. II.2 we compare explicitly the effective sky coverages as a function of  the X-ray flux 
limit for the three most representative present surveys and the WAXS Shallow and Deep surveys.   
Note how in both case, the WAXS surveys represent a step of an order of magnitude with respect 
to existing surveys.   Note also that these flux limits are conservative, as they have been based on 
the request of collecting 50-100 counts from a cluster at the limit.   In practice, we can very 
probably push our detection limit down by a factor of two in both surveys.

Based on the experience gained with the RDCS survey (Rosati et al. 1998), and the similar survey 
by Vikhlinin et al. (1998b), we define as ``typical" a cluster with an extension of $\sim 1$ arcmin 
radius. This represents the median angular size (roughly twice the 50\% power radius) of the 
clusters in the RDCS sample after de-convolving the ROSAT-PSPC PSF.

In Fig. II.3, we plot the limiting flux as a function of the exposure time, for a signal-to-noise ratio 
(S/N)=5 and for a source with an extension of 1 arcmin radius. For a cluster described by a 
Raymond-Smith thermal model, with KT=5keV and 0.3 solar abundance (filtered by a galactic 
absorbing column density equal to $3\times10^{20}$ cm$^{-2}$), the conversion factor is $1 cts s^{-1} = 1.2\times 10^{-11}$ erg s$^{-1}$ 
cm$^{-2}$ (between 0.5-2.0 keV), which is accurate within 10\% for clusters with temperatures between 
2 and 10 keV.   We considered an instrumental and cosmic background of $10^{-3}$ cts s$^{-1}$ arcmin$^{-2}$, 
which is about two times the diffuse X-ray background in the WFXT energy band and should 
provide a conservative upper limit.  In Fig. II.3 we also show the S/N sensitivity curve in the case 
of a total background two times lower and the $S/N=10$ and $S/N=50$ sensitivity curves.

If we take an exposure time of  $>10^5$ s for the deep area (100 sq.deg.) and $>10^4$ s for the shallow 
area (900 sq.deg.), from Fig. II.3  we can estimate a limiting flux (0.5 - 2.0 keV) for our ``typical 
cluster" of $\sim 10^{-14}$ ergs cm$^{-2}$ s$^{-1}$ and $\sim 5\times 10^{-14}$ erg cm$^{-2}$ s$^{-1}$ 
for the deep and shallow area, 
respectively. These exposure times ensure that, at the faintest fluxes here considered, at least 50 
to 100 net counts will be accumulated from a typical cluster. {\bf Based on the RDCS survey, the 
corresponding signal-to-noise is enough to discriminate between a point-like and an 
extended source, thus allowing a robust list of cluster candidates to be defined by using X-ray 
data alone}. This clearly implies that we shall also be able to detect clusters at fainter fluxes. 

{\bf A very important cosmological probe is provided by the evolution of clusters}.

The main reason for this is that clusters correspond to the high peaks of the primordial density field 
(e.g. Kaiser 1984), so that their abundance (i.e. the number of clusters within a given mass range), is 
highly sensitive to the details of the underlying mass density field.   The typical mass of rich clusters 
($\sim 10^{15}\msole$), is close to the average mass within a sphere of  16 Mpc radius in an unperturbed Universe, 
so that the local ($z<0.2$) abundance of clusters is expected to place a constraint on the r.m.s. 
fluctuations on the same scale, what is called $\sigma_8$ (being  normally expressed using $H_0=100$ km s$^{-1}$ Mpc$^{-1}$).   

This is basically a measure of the normalization of the power spectrum, and the Press-Schechter (1974) 
theory, easily shows that the cluster abundance is highly sensitive to $\sigma_8$ for a given value of the density 
parameter $\Omega_0$ (see Borgani et al. 1998 for details).  At the same time, once the local abundance of 
clusters (i.e. $\sigma_8$) is fixed, its evolution back in time will mainly depend on  $\Omega_0$.   
{\bf Therefore, if we are 
able to trace the cluster abundance to high redshifts in a reliable way, we shall directly constrain 
the value of the cosmological density parameter}.

The problem is that we cannot observe directly the abundance of clusters within a defined mass range, 
as required by the theory, and have to resort to some kind of observable which can be connected as 
closely as possible to mass.  Cluster masses can be measured through galaxy velocity dispersions, but 
this is very time-consuming and for the moment limited to local samples.  Weak-lensing maps are also 
a promising technique, but again it is difficult to collect systematic observations for large samples.  {\bf In 
this respect, X-ray selected clusters offer the best opportunity, as they have measurable 
properties that can be linked to mass in a more direct way than optically-selected systems}.  The 
best parameter would be X-ray temperature, which offers the most direct route to mass. Although the 
situation is improving (Henry 1997), X-ray temperatures are however still not available in a systematic 
way for large samples of clusters.   An easier way is to use X-ray luminosities.  Considerable efforts 
have therefore concentrated in the last few years on trying to detect signs of evolution with redshift in 
the {\bf X-ray luminosity function} (XLF), to be then related to the mass function through the 
Luminosity-Temperature relation.   

After the pioneering results from EMSS survey (Gioia et al. 1990), in the last couple of years there has 
been a major burst of works tackling this problem, all based on serendipitous searches of clusters over 
deep ROSAT pointed observations from the public archive (Rosati et al. 1998, Vikhlinin et al. 1998b; 
see Rosati 1998 for a review).  These studies, that were able to reach fluxes as faint as $2\times 10^{-14}$ erg s$^{-1}$ cm$^{-2}$,
have shown how there is practically no evolution between $z=0$ and $z=0.8$ in the abundance of clusters 
of moderate luminosity ($L < L_*\sim 4\times 10^{44}$ erg s$^{-1}$), while there is a hint for evolution in the abundance of 
very luminous systems.    Vikhlinin et al. (private comm.), in fact find no luminous cluster in their 
distant redshift bin ($z>0.5$), while 9 would be predicted on the basis of the local XLF. Fig. II.4 shows a 
comparison of the XLF in local and distant samples.  Note how in this figure, prior to the latest 
Vikhlinin et al. result,  the only evidence for evolution is the $\sim 2\,\sigma$ deficiency in the EMSS XLF at high $z$ 
(starred symbols).  

These serendipitous surveys are in general limited by the difficult compromise between depth 
(i.e. flux limit) and sky coverage, as we further show in Fig. II.5 for the RDCS: the data cover 
different ranges of luminosity at different redshifts and direct comparison is difficult.  {\bf The only 
way to improve significantly on these measures of evolution is to enlarge the covered area at 
similar fluxes, and to go fainter on comparable areas.   This is exactly what the Shallow and 
Deep WAXS surveys will do. Determining the existence of luminous clusters at $z>1$, 
providing a robust measure of the XLF at different redshifts and establishing firmly the 
evolution of clusters of galaxies is a primary goal of the mission}.

{\bf The full power of the survey concerning the statistics of large-scale structure will be 
exploited through the measurements of redshifts for WAXS groups and clusters}.   While 
``local" ($z<0.2$) surveys of rich clusters, as the REFLEX survey, give a coarse view of the large-
scale distribution of matter, the WAXS survey will be unique in producing  a large sample of 
groups at $z<0.1$, that will give a more detailed description of local large-scale structure.  Even 
more important, this will be based on objects selected through a clean tracer of mass as X-ray 
emission is. The Shallow survey will detect nearly 500 groups within $z<0.1$.

A direct example of the power of using X-ray selected clusters for studying the large-scale structure of 
the Universe has been recently provided by the ROSAT-ESO Flux Limited X-ray cluster survey 
(REFLEX, Guzzo et al. 1995, Boehringer et al. 1998).  While the above mentioned ROSAT deep 
surveys are all based on searches of serendipitous cluster sources on archival pointed observations (see 
Rosati 1998 for a review), the REFLEX survey is a wide-angle project based on the ROSAT All-Sky 
Survey (RASS), which is about two orders of magnitude brighter in flux limit ($\sim1 \times 10^{-12}$ 
erg s$^{-1}$ cm$^{-2}$ for clusters).    The preliminary results from REFLEX represent a very good example of the 
effectiveness of X-ray selected clusters as tracers of large-scale structure.  In Fig. II.6 we plot the 
estimate of the power spectrum from a preliminary version of the REFLEX data.    This is computed 
using only 230 clusters with $z<0.1$, while the whole survey is going  to contain 475 clusters with $z<0.3$ 
and flux brighter than $3\times  10^{-12}$ erg s$^{-1}$ cm$^{-2}$, over an area of 4.24 sr.   

{\bf The WAXS surveys will be able to both increase the detail of this measure in the local ($z<0.2$) 
Universe, by using X-ray selected groups and poor clusters, and, most importantly, to perform 
the same measure as a function of redshift, out to $z\sim1$, using X-ray luminous rich clusters}.  This 
will represent an unprecedented probe of the evolution of the large-scale structure of the Universe, 
constraining the cosmological parameters.

{\bf Very little is known about the physics of the intercluster medium in superclusters}, primarily 
because these systems are rare (only 3 superclusters are known with masses comparable to Shapley 8) 
and because they are very hard to map due to their large angular size. Possible detection of a significant 
number of rich superclusters in the WAXS surveys provides a unique opportunity for their detailed 
study.  A 30 Mpc diameter supercluster will subtend approximately $1\deg$ at $z=0.5$.  With the large 
contiguous survey area of the WAXS surveys, available only to this specific mission, we can readily 
map the X-ray background and provide strong limits on any diffuse emission within the core of  such a 
supercluster. A very conservative estimate of the minimum detectable surface brightness enhancement 
is 30\% of the X-ray background brightness around 1 keV. For a 0.5-1 keV plasma filling a 10 Mpc 
supercluster, this corresponds to the central density of $1-2\times 10^{-5}$ cm$^{-3}$. The diffuse intercluster gas in 
these superstructures can be detected in the WAXS surveys out to substantial redshifts. According to 
our visibility simulations, the diffuse gas in a Shapley-like supercluster  would be detected out to 
$z\sim 0.5$ in the Deep survey and out to $z\sim0.2$ in the Shallow survey.   While other X-ray missions (e.g., 
XMM) may detect comparable numbers of clusters and superclusters, they will be randomly spread 
over the sky and the primary science objective -- studying large scale structure -- is feasible only with 
the large contiguous surveys proposed here.

A sensitive X-ray survey like WAXS should also be able to map directly the warm/hot gas present 
between clusters, trapped inside the potential filaments and superstructures.   In fact, at redshifts below 
0.5 also the diffuse gas filling the deepest parts of the supercluster and filament potential wells is 
starting to be detected in the X-rays (see, e.g., Wang et al. 1997, Connolly et al. 1996).   This allows to 
map directly the large scale distribution of the warm/hot baryons, the dominant baryonic component of 
the matter in the Universe (Cen \& Ostriker 1998). The volume fraction of the warm/hot gas is only  a 
few  percent while the relative mass fraction is up to $\sim 50\%$ at $z>0.5$ (see Fig. 2 in Cen \& Ostriker 1998) 
indicating that the filamentary Cosmic Web is the repository of such abundant baryonic material. At the 
same redshifts the overall emission of the warm/hot gas with $k\,T > 0.5$ keV is just a factor  $\sim 3$ below the 
overall emission of the hot gas in rich clusters (see Fig. 4 in Cen  \& Ostriker) indicating that such 
warm/hot gas is a major contributor to the soft ($k\,T > 1$  keV) diffuse X-ray background.  Also, Colberg 
et al. (1998) showed that clusters accrete matter from a few preferred directions, defined by the 
filamentary structures, and that the accretion persists over cosmological long times.   A spectacular 
example is shown in Fig. II.7.
 
{\bf The observational study of AGNs, along with the theoretical study of their formation, is another 
avenue to better understanding the origin of structures}.
 
AGNs dominate the deep X-ray images, comprising approximately 80\% of all the sources (Hasinger et 
al. 1998; Schmidt et al. 1998) at high galactic latitude.  Once the clusters (and the bright stars) have 
been identified, one can reasonably assume that most of the remaining sources are AGNs.  In the 
present baseline for the two high latitude WAXS surveys  ($\sim 100$ sq.deg. at  $S_X > 4 \times 10^{-15}$ erg s$^{-1}$ cm$^{-2}$ 
and  $\sim 900$ sq.deg. at $S_X > 3 \times 10^{-14}$ erg s$^{-1}$ cm$^{-2}$  at $5\,\sigma$) we expect to detect $\sim 30,000 - 40,000$ 
AGNs, approximately equally divided in the two surveys.  These numbers will increase by about 50\% 
pushing the detection limit down to $4\,\sigma$. While most of these AGNS will be the ``classical" broad-line 
quasars, a not negligible fraction is expected to be constituted by absorbed AGNs, with $N_H > 10^{21}$ cm$^{-2}$ 
(Comastri et al. 1995).

At the $5\,\sigma$ limiting fluxes quoted above, we will detect $\sim 15$ and $\sim 200$ AGNs per square degree in the 
Shallow and Deep surveys, respectively. The surface density of AGNs in the Deep Survey is 
significantly higher than those which will be obtained in the forthcoming optical surveys on large areas. 
For example, the Sloan Digital Sky Survey (SDSS) will measure redshifts of $\sim 105$ QSO in $10^4$  sq.deg.  
(Margon 1998), while the expected surface density of AGNs in the 2dF sample is 30-40 sq.deg. It 
should be noted that the optical surveys will select magnitude-limited samples of AGN with colors as a 
primary selection criterion. Since in the optical band AGNs are a small fraction of the total number of 
objects, the statistical uncertainties in the colors at faint magnitudes and the difficulty in separating the 
point-like objects from the much more numerous faint, extended galaxies makes very difficult the 
assessment of the level of completeness of faint optically selected samples. From this point of view, the 
X-ray selection is highly superior because a very large fraction ($> 70\%$) of X-ray sources are known to 
be AGNs.

The comoving spatial density of AGNs detected in WAXS surveys will be $\sim 10^{-5}$ Mpc$^{-3}$, i.e. 
comparable to that of galaxy groups at low redshift. Thus, AGNs are promising tracers of the large 
scale structure, although they cannot be substituted to clusters for a quantitative measurement of the 
matter power spectrum. The physical cause of AGN activity is not well understood yet and therefore 
AGNs can be arbitrarily biased (or anti-biased) with respect to the total matter distribution. 
Nevertheless, AGNs should be suitable for qualitatively mapping the large scale structure. For this 
purpose, the high spatial density of X-ray selected QSOs and our square survey geometry is highly 
desirable.

The evolution with redshift of the clustering strength is much more controversial, with contradictory 
results obtained by different groups using samples of $\sim 1000$ quasars spanning the entire redshift range 
(see, for example, Andreani \& Cristiani 1992).  These results will be soon improved by the analysis of 
the forthcoming 2dF quasar sample, which ultimately will contain about 30,000 quasars with $m_B < 21.0$ 
in 750 sq.deg. Even this sample, however, will not provide much information for $z>2.5$, because of the 
relatively bright limiting magnitude and, therefore, the relatively small number of objects at such high 
redshifts. 

Vice versa, WAXS will allow the detection of a large number of quasars at $z>2.5$. A precise estimate 
of how many such quasars will be detected is highly uncertain because very little is known about the X-
ray luminosity function (XLF) at these redshifts. While most of the AGNs will be in the range $0.5<z<2.5$, 
more than 2,000 quasars (most of them in the deep survey) are expected to be detected at $z>2.5$.  
More pessimistic models, with a decreasing comoving density beyond $z=2.5$, still predict about 1,000 
such objects. These estimates are consistent with the available optical identifications in the deep 
ROSAT survey in the Lockman Hole (Schmidt et al.  1998). Note that in this survey, covering only  
$\sim 0.2$ sq.deg., it has been detected the highest redshift, X-ray selected quasar ($z=4.45$; Schneider et al. 
1998), with a flux higher than the limit of the WAXS Deep survey. 

{\bf WAXS is an excellent and unique mission for such a project}.  The ROSAT all-sky survey is too 
shallow and the ROSAT deep surveys have too small solid angles. Serendipitous surveys with AXAF 
and XMM eventually will cover a large area at a limiting flux similar to that of the WAXS. The 
disadvantage of a serendipitous survey, however, is that it will be spread over of thousands pointings in 
different directions, which complicates the optical follow-up and makes the detailed study of the AGN 
spatial distribution impossible.

{\bf Rapid and large amplitude variability is common among AGNs}, both for radio-quiet Seyfert 
galaxies and for radio-loud objects, and it is the main defining property of blazars. For the shallow 
survey, each field of view will be observed in a single passage for approximately $10^4$ seconds, while for 
the small area deep survey each field will be observed $\sim  10$ times, within a total period of a few months. 
In this area we will therefore have the possibility to detect variability on both short timescales (of the 
order of hours) and long timescales (from a few weeks to a few months).  The latter timescales are 
particularly interesting, since they have not been well sampled yet, except for a few selected sources. 
Time variability studies on these scales will be impossible with other missions (with the exception of a 
small number of ``famous" sources). The number of AGNs subject to this variability analysis is huge 
(essentially, the brightest 30\% of all objects), allowing us to define for the first time the variability 
properties of all classes of AGNs.

{\bf We will also include in the mission plan, a Galactic Plane survey to better understand the 
structure of the Milky Way and its X-ray properties.}
 
Given the strong dependence of coronal activity on rotation and age, any flux limited X-ray survey will 
preferentially detect active stars which are observable to larger distances than low-activity ones.  This 
explains why flux limited surveys carried out with Einstein, EXOSAT and  ROSAT have typically 
shown a large fraction of active stars, either young stars or RS CVn binaries (Favata et al. 1993, 1995; 
Tagliaferri et al. 1994; note that for RS CVn binaries the high rotation and high coronal activity is due 
to tidal interaction rather than young age). X-ray observations thus provide a unique way to investigate 
the distribution of active stars, and in particular of young stars, in the Galaxy up to distances of few 
kpc.

There are several factors that are expected to influence the distribution of X-ray active stars in the solar 
neighborhood, all of which are still poorly understood. A first dominant component is expected to arise 
from the structure of the galactic disk, with young stars strongly concentrated on the galactic plane and 
rapidly decreasing at higher galactic latitudes.  To first approximation, this distribution should be only 
weakly dependent on galactic longitude, provided the sampled volume remains sufficiently close to the 
Sun (up to few hundred parsec, as in most X-ray surveys carried out so far). However, as soon as we 
increase the sensitivity, we will start exploring larger and larger volumes and the radial distribution of 
stars on the disk (e.g. the spiral arm structure) will become increasingly important. For sufficiently high 
sensitivity, a clear asymmetry between the directions of the galactic center and anticenter should 
become readily apparent. Even at limited sensitivity, the distribution of young active stars should be 
markedly different at different galactic latitudes, owing to the finite scale-height ($\sim 100$ pc) of their 
density perpendicular to the galactic plane.
 
A cross-correlation of the RASS survey (at a flux limit of  $\sim2 \times 10^{-13}$ erg s$^{-1}$ cm$^{-2}$) with the Tycho 
catalogue (which is complete down to $m_V =10.5$) has recently shown an additional structure in the 
spatial distribution of X-ray active stars besides the general decrease with galactic latitude (Guillout et 
al. 1998a, b; see Fig. II.8). This density enhancement, which is particular prominent in the third and 
fourth quadrants of galactic longitudes (i.e. between $l = 180\deg$ and $l = 360\deg$), is in very good agreement 
with the expected position of the so-called Gould Belt (GB). This is a large-scale ring structure of 
recent star formation which had previously been identified on the basis of the spatial distribution of OB 
associations and which appears to be inclined by about $20\deg$ to the galactic plane. Fig. II.9 shows a 
model prediction at a sensitivity a factor 4 higher and for stars down to $m_V = 15$ (simulation courtesy of 
P. Guillout); the galactic plane structure and the GB are now detectable much more clearly.  

For a few selected regions at low galactic latitudes in Cygnus and in Taurus covering respectively 64.5 
sq.deg and 70 sq. deg, a complete optical identification program of the RASS sources has been carried 
out (Motch et al. 1997), showing that $\sim 85\%$ of the RASS sources at low galactic latitudes are indeed 
stars. A GPS at a sensitivity of  $\sim 1-2 \times 10^{-14}$ erg sec$^{-1}$ cm$^{-2}$ has been carried out (cf. Morley at al. 1996, 
Pye et al. 1997, Sciortino et al. 1998) using a number of individual PSPC pointings in the range of 
galactic longitudes from $l = 180\deg$ to $270\deg$ and at very low galactic latitudes $\| b \| < 0.3\deg$ . The sensitivity 
is much larger than that of the RASS but the total survey area is only 2.5 sq.deg and makes the results 
statistically uncertain and dependent on local fluctuations.

With the current scan rate of 0.3 arcsec/sec and a FOV of $1\deg \times 1\deg$, current estimates of the WFXT 
sensitivity indicates that a limiting sensitivity of  $\sim 2 \times 10^{-14}$ erg s$^{-1}$ cm$^{-2}$ (appropriate for moderately 
absorbed - $N_H < 10^{21}$ cm$^{-2}$  - thermal sources with a temperature of  $\sim 10^7$ K) in the spectral band 
0.35 - 8 keV can be reached for point sources with a single scan (i.e. for an exposure time of 10 ksec).  
With the goal to survey $\sim 500$ sq.deg. at this limiting sensitivity and assuming a 15\% overlap between 
adjacent strips we estimate that the overall observing time will be of $7\times10^6$ s equivalent, for a 65\% 
average observing efficiency, to $\sim 4$ months of elapsed time to be devoted by WFXT to the GPS.

Hence, at the same limiting sensitivity, WFXT will cover an area 200 times larger than the ROSAT 
GPS based on pointed observations.  With respect to the areas studied in detail by Motch et al. (1997) 
in Cygnus and Taurus, the improvement in area coverage by WFXT will be only a factor 3.7, but the 
sensitivity will be an order of magnitude higher, making accessible a volume 100 times larger than with 
the RASS.  In all these cases, the step forward with respect to previous observations allowed by the 
proposal WFXT GPS will be enormous.

{\bf An ultra-deep survey of a few square degrees has been included in the mission plan}, for detecting 
the faintest possible sources before being limited by source confusion. This survey requires that we 
reach the goal of better than 15-arcsecond angular resolution for the optics. Should this not be the case, 
the ultra-deep survey will not be made, and we will instead extend the area of the shallow and deep 
surveys. 

We have simulated a very deep WFXT image using the BeppoSAX simulator, assuming an Half 
Energy Width of 12 arcsec over the whole field of view ($1\deg$) and to this we have applied the Waveket 
Transform algorithm already developed at OAB.  At $10^6$ seconds integration time (see Fig. II.10) we 
are getting close to the confusion limit. However, we still see the sources quite well. In Fig. II.11 we 
show the comparison between the input sources and those detected by the algorithm. Although we did 
not spend to much time in refining the algorithm to the WAXS case, the agreement is excellent. We can 
recover all sources down to a flux limit of $<6\times10^{-16}$  erg s$^{-1}$ cm$^{-2}$. Moreover, our simulations have also 
shown that we are accurately able to distinguish between point-like and extended objects for sources 
with only 50 counts. 

\vskip .2in
\section{The Instrument}

The configuration of the satellite as shown on Fig. III.1 is the result of the trade-off and detailed 
configuration design and analysis.

The WFXT is the assembly of several structural and functional elements, in the following we describe 
the mirror module and illustrate the overall response. The Mirror Support Adapter (MSA) is the 
outermost element, its main function is to support WFXT connecting the telescope to the satellite 
structure. The next element, moving towards the core is the case: it connects the top of the MSA to the 
front spider. The front spider is the element supporting the mirror shells. In the front spider two circular 
rings, ``C" shaped, are connected by 16 radial spokes. The spokes have a rectangular cross section; 
moving from the inner ring to the outer ring, their height decreases linearly while their width increases. 
Underneath the front spider are placed two flat masks (X-ray pre-collimator). They are connected to the 
case through a circular ``C" shaped ring - X-ray pre-collimator support. On the other side, the front 
spider supports 25 mirror shells (MSs). The connection between the mirror shells and the front spider is 
by means of glue. The 25 MSs are concentrically disposed, they are divided into two groups: the 
outermost ones, formed by the MSs from \#1 to \#9, while the innermost range from \#10 to \#25. The 
front spider supports also a cylindrical element: the fiducial light mechanical interface. The MSA top is 
connected to the rear spider. The rear spider is the assembly of two circular rings, having a rectangular 
cross section, connected by means of 16 spokes. The main function of these spokes is to support the 
thermal baffles and the electron diverter. The thermal baffles are three concentric cylindrical shells.

The final mirror module consist of the MSS, thermal pre- and post-collimators and an X-ray pre-
collimator (see Fig. III.2). 

In the Phase A study we considered in detail three models for the WFXT Mirror Module based on 
different technologies: 

$\bullet$ Mirror Module with Nickel Mirror Shells (named Model A);

$\bullet$ Mirror Module with SiC Mirror Shells (named Model B);

$\bullet$ Mirror Module with Al2O3 Mirror Shells (named Model C).

All the models have the same interfaces with respect to the tube of the satellite and the fiducial light 
system. The design of each model has been done taking into account:

$\bullet$ The optical specifications;

$\bullet$ The interface requirements;

$\bullet$ The experience that Media Lario has accumulated as responsible for the mirror module 
design and manufacturing of other X-ray astronomy projects (JET-X, XMM). 

For each Model we performed a Finite Element Model (FEM) and a thermo-structural analysis (with 
the impact on the optical performance of the Mirror Module).
 
A prototype SiC carrier has been manufactured by Morton International (USA) and C. Zeiss (D) has 
made the replica. This corresponds to the largest mirror shell of the WFXT telescope. This shell has 
been tested at the X-ray PANTER facility. Measurements were carried out with the PSPC and CCD 
detectors at 0.5 and 1.5 keV. The results of these tests are very encouraging; we have obtained values of 
the $HEW \leq 15"$ (Fig. III.3). {\bf The breakthrough is that we have fully demonstrated that by 
adopting the polynomial design it is possible to have almost constant HEW on a large field of 
view ($\pm 30'$). This is an outstanding result if compared with the performances of , e.g., the JET-X 
Wolter I design}.

We point out, however, that the HEW values are higher than what we would expect from the design 
and manufacturing errors. We have identified an epoxy variation thickness at the front and back 
entrance of the mirror shell, which has caused a variation in the mirror profile and in turn the image 
blur. This problem can easily be solved and a new mirror shell is under manufacturing to confirm our 
analysis.
 
In particular, extra- and infra-focal position PSPC images show that the image degradation took place 
in a limited azimuth sector of the mirror shell. By extrapolating the performances of the azimuth 
portion of the mirror not degraded by the epoxy variation thickness we are able to derive an 
approximate value of the best HEW that can be obtained once solved this problem. The procedure 
involves the de-convolution of the PSPC resolution and provides a HEW of $\leq 12$ arcsec. {\bf This 
demonstrates that the mirror technology fabrication meets the requirement for the WFXT 
program}. 

The collecting area of the telescope at 1.5 keV is $\sim 360$ cm$^2$ ($\sim 310$ cm$^2$ after convoluting with filter and 
CCD response).  The energy resolution, at 1.5 keV, is $\Delta E/E \leq 10\%$. The CCD focal plane is based on 
the heritage of the technology developed for JET-X, XMM and AXAF using the best technology and 
detectors available today. The convolution of the mirror shells with the filter and detector shows that 
we have a rather good response in the range of interest (0.2 - 8 keV). The response curve of the 
instrument is plotted in Fig. III.4. The high angular resolution distinguishes extended clusters of 
galaxies from point-like sources at any redshift, and the good sensitivity reaches a large number of very 
faint and distant objects as required to achieve the science goals of the mission.

\vskip .2in
\section{A simulated mission}

A typical mission plan has been computed for the two possible orbital inclinations: 3.5 and $50\deg$. The 
purpose here, is to show that the scientific goals can really be accomplished in the minimum mission 
baseline of approximately two years. The proposed mission plans are computed with the purpose to 
keep the amount of large satellite movements (e.g. switching among different targets) as low as 
possible. All factors reducing the observing efficiency  (Earth and solar interference, radiation belt 
crossing, etc.) have been taken into account. For simplicity in the table we list only the simulation for 
the 3.5 degrees orbit. Note that in the last column, total, we give the percentual accumulation time achieved 
after each pointing to reach the full integration time at the specified target, 100\%.

{\bf The data will be promptly released after completion of the observations and in any case within 
one year of the end of the mission. They will enter the public domain through the delivery of the 
calibrated data to the appropriate centers. The final calibration will be implemented and 
released at the end of the mission when all LMC data will be available. Copy of the archives will 
be set up in Italy, USA, UK and Germany and will be regulated by common rules}. The clean, 
calibrated data will be kept available online on disks; at the end of the mission this service could be 
under the responsibility of the ASI-SDC. When justified a large set of data could be transferred to the 
requesting group by the most convenient means of data transfer (DAT tapes, CD-ROMs). {\bf The WFXT 
Science team, coordinated by the PI, will release at the end of the mission the final and official 
catalog of the detected sources, along with information such us coordinates, fluxes, source 
extension, etc}.

\vskip .2in
\section {Acknowledgements}

The Scientific Institutes involved in the hardware aspect during the Phase A study are: Brera 
Astronomical Observatory (OAB), Smithsonian Astrophysical Observatory (SAO), University of 
Leicester (UL), Max-Planck-Institute f\"ur Extraterristrische Physik (MPE). The Industries 
participating are: ALENIA, MEDIA-LARIO, TELESPAZIO, OFFICINE GALILEO.  P.I: G. 
Chincarini (OAB); co-PI:S. Murray (SAO); Co-I J. Tr\"umper (MPE), A. Wells (UL), Telescope 
PI: O. Citterio (OAB); PM: G. Tagliaferri (OAB); PS: S. Sciortino (OAPA).
Six panels headed by S. Sciortino coordinated all the work for the science proposal. The panels 
were chaired by: Colafrancesco (LSS), Chincarini (Clusters of galaxies), Zamorani (AGN), 
Forman (Galaxies) Pallavicini (Stars) and Watson (Compact Objects).  Contribution to this work 
came from Antonuccio-Delogu, Arnaboldi, Bandiera, Bardelli, Boehringer, Bonometto, Borgani, 
Campana, Catalano, Cavaliere, Covino, Della Ceca, De Grandi, De Martino, Fiore, Garilli, 
Ghisellini, Giacconi, Giommi, Girardi, Giuricin, Governato, Guillout, Guzzo (who revised the 
final version of the science proposal), Iovino, Israel, Lazzati, Le Fevre, Longo, Maccacaro, 
Maccagni, Mardirossian, Matarrese, Micela, Molendi, Molinari, Moscardini, Murray, Norman, 
Perola, Osborne, Pye, Ramella, Randich, Robba, Rosati, Scaramella, Stella, Stewart,  Tagliaferri, 
Tozzi, Trinchieri, Vikhlinin, Vittorio, Wolter.
Members of the following Institution expressed direct interest in the mission: ASI (SDC), 
Bologna (OA), Catania (OA), Firenze (OA), Milano (IFCTR,UNI), Napoli (OA), Padova 
(UNI,OA), Palermo (OA,UNI), Perugia (INFN), Roma (OA-UNI), Trieste (OA), Marsiglia 
(CNRS,LAS), Baltimore (IHU), Cambridge (MIT), Princeton (UNI), Munich (ESO,MPE), 
Copenhagen (DSRI).

Indeed this paper is the result of cutting and pasting part of the Phase A proposal.

\vskip .2in
\section{References}

\rref Allen, S.W., Fabian, A.C., 1998, MNRAS, 297, L63
\rref Andreani, P.,  Cristiani, S., 1992, ApJL, 398, L13
\rref Bahcall, N. A., 1979, ApJ, 232, 689
\rref Bardeen et al., 1986, ApJ, 304, 15
\rref Bardelli S., et al., 1997, Astroph. Lett. \& Comm., 36, 251
\rref Bode N., et al. 1998, preprint 
\rref Boehringer, H., Guzzo, L., Collins, C.A., Neumann, D.M., Schindler, S., et al. (the REFLEX Team), 
1998, ESO Messenger, 94, in press (astro-ph/9809382)
\rref Borgani, S., Rosati, P., Tozzi, P., \& Norman, C., 1998, ApJ, in press
\rref Branduardi-Raymont G., et al. 1994, MNRAS 270, 947
\rref Briel, U., Henry, P., 1995, A\&A, 302, 9
\rref Bryan, G., 1997, PhD Thesis
\rref Carroll, S. M., Press, W. H., Turner, E. L., 1992, ARA\&A, 30, 499 
\rref Cen, R., Ostriker, J. P., 1998, Science, in press (astro-ph/9806281) 
\rref Colberg, J.M., White, S.D.M., Jenkins, A., Pearce, F.R., 1997, submitted to MNRAS, astro-ph/9711040
\rref Collins, C. A., Burke, D. J., Romer, A. K., Sharples, R.  M., Nichol, R. C., 1997, ApJ, 479, L117
\rref Connoly A. J., et al., 1998, ApJ, 473, L67
\rref Comastri, A., Setti, G., Zamorani, G., Hasinger, G., 1995, A\&A, 296, 1
\rref Dalton, G. B., Croft, R. A. C., Efstathiou, G., Sutherland, W. J., Maddox, S. J., Davis, M., 1994, 
MNRAS, 271, L47
\rref David, L. P., Jones, C., and Forman, W., 1995, ApJ, 445, 578
\rref De Grandi, S., Guzzo, L., Boehringer, H., Molendi, S., Chincarini, G., 1998, ApJ, submitted
\rref Donnelly, R. H., Faber, S., O'Connell, R., 1990, ApJ, 354, 52
\rref Durret, F., Forman, W., Gerbal, D., Jones, C., Vikhlinin, A.,  1998, A\&A, 335, 41
\rref Ebeling, H., Edge, A. C., Fabian, A. C., Allen, S. W., Crawford, C. S.,  Bohringer, H., 1997, 
ApJ, 479, L101
\rref Efstathiou, G., 1995, MNRAS, 272, L25
\rref Eke, V. R., Cole, S., Frenk, C. S., 1996, MNRAS, 282, 263
\rref Elvis, M., et al., 1981, ApJ, 246, 20 
\rref Ettori, S., Fabian, A. C., White, D. A., 1998, MNRAS. 289, 787
\rref Fabbiano, G., Trinchieri, G., 1985, ApJ, 296, 430
\rref Favata F., Barbera, M., Micela, G., Sciortino, S., 1993, A\&A 277, 428
\rref Favata F., Barbera, M., Micela, G., Sciortino, S., 1995, A\&A 295, 147
\rref Favata F., Micela G, Sciortino S., Vaiana G. S., 1992, A\&A, 272, 124 
\rref Forman, W., Jones, C. \& Tucker, W., 1985, 293, 102
\rref Fukugita, M., Hogan, C.J. and Peebles, P.J.E., 1997, astro-ph/9712020
\rref Geller, M. J., Huchra, J. P., 1989, Science, 246, 897
\rref Gioia, I. M, Maccacaro, T.,  Schild, R. E., Wolter, A., Stocke, J. T., Morris, S.L.,
Henry, J.P., 1995, ApJS, 72, 567
\rref Griffiths, R. E., Georgantopoulos, I., Boyle, B. J., Stewart, G. C., Shanks, T.,
Della Ceca, R, 1995, MNRAS, 275, 77
\rref Groth, E. J., Peebles, P. J. E., 1977, ApJ, 217, 385
\rref Guillout P., Haywood, M., Motch, C., Robin, A.C., 1996, A\&A 316, 89
\rref Guillout P., Sterzik, M. F., Schmitt, J. H. M. M., Motch, C., Egret, D., Voges, W., 
Neuh\"auser, R., 1998a, A\&A, in press
\rref Guillout P., Sterzik, M. F., Schmitt, J. H. M. M., Motch, C., Neuh\"auser, R., 1998b, 
A\&A, in press
\rref Guzzo, L., Boehringer, H., Briel, U., Chincarini, G., Collins, C. A., et al., 1995, in ``Widerref Field 
Spectroscopy and the Distant Universe", S.J. Maddox \& A. Aragonrref Salamanca, eds., World 
Scientific, Singapore, p.205
\rref Guzzo, L., Strauss, M. A., Fisher, K.B., Giovanelli, R., \& Haynes, M.P., 1997, ApJ, 489, 37
\rref van Haarlem, M. P., Frenk, C. S., White, S. D. M., 1997, MNRAS, 287, 817
\rref Hanami H. et al. 1998, preprint
\rref Hasinger, G., Burg, R., Giacconi, R., Hartner, G., Schmidt, M., Tr\"umper, J., Zamorani, G., 1993, 
A\&A 275, 1
\rref Hasinger, G., Burg, R., Giacconi, R., Schmidt, M., Trumper, J., Zamorani, G., 1998, A\&A, 329, 482
\rref Hauser, M. G. \& Peebles, P. J. E., 1973, ApJ, 185, 757
\rref Henry, J. P., 1997, ApJ, 489, L1
\rref Henry, J. P., Gioia, I. M., Maccacaro, T., Morris, S. L., Stocke, J. T, Wolter, A., 1992, ApJ, 386, 
408
\rref Hjorth, J., Oukbir, J., van Kampen, E. 1998, MNRAS, in press (astro-ph/9802293)
\rref Hudson, M. J., Ebeling, H. 1997, ApJ, 479, 621
\rref Jenkins, A., et al. 1998, ApJ, 499, 20
\rref Kaiser, N., et al., 1998, ApJ, submitted (astro-ph/ 9809268)
\rref Landy, S. D., Szalay, A. S. 1993, ApJ, 412, 64
\rref Long K. S., Helfand D. J., Grabelsky D., 1981, ApJ, 248, 925
\rref Maccacaro, T.,  Schild, R. E., Wolter, A., Henry, J. P., 1991, ApJS, 76, 813
\rref Maddox, S. J., Efxtathiou, G., Sutherland, W. J., and Loveday, J., 1990, MNRAS, 242, 43p
\rref Micela, G., Sciortino, S.,  Serio, S., Vaiana, G. S., Bookbinder, J., Golub, L., Harnden, F. R. Jr.,
Rosner, R., 1985, ApJ, 292, 172
\rref Micela, G., Sciortino, S.,  Vaiana, G. S., Schmitt, J. H. M. M.,  Stern, R., Harnden, F. R. Jr.,  
Rosner,  R., 1988, ApJ, 325, 798 
\rref Micela, G., Sciortino, S., Vaiana, G. S., Harnden, F. R. Jr., Rosner, R., Schmitt, J.H.M.M., 
1990, 348, 557
\rref Micela, G., Sciortino, S., Favata, F., 1993, ApJ 412, 618
\rref Mo, H. J., Jing, Y. P., White, S. D. M., 1996, MNRAS, 284, 189
\rref Morley, J. E., Pye, J. P., Warwick, R. S., Pilkington, J., 1996, MPE Report 263, 659
\rref Moscardin,i L., et al., 1998, preprint
\rref Motch, C., Guillout, P., Haberl, F., Pakull, M., Pietsch, W., Reinsch, K., 1997, A\&A 318, 111
\rref Neuh\"auser, R., 1997, Science 276, 1363
\rref Ostriker, J. P., Cen, R., 1998, submitted to Science,  astro-ph/9806281
\rref Ostriker, J. P., Steinhardt, P. J. 1995, Nature, 377, 600
\rref Page, M. J., Mason, K. O., McHardy, I. M. et al., 1997, MNRAS, 291, 324
\rref Pallavicini, R.,  Golub, L., Rosner, R., Vaiana, G. S., Ayres, T., Linsky,  J.L., 1981, 
ApJ, 248, 279
\rref Pallavicini, R., 1989, A\&A Review,  1, 177
\rref Pallavicini, R., 1998, Space Science Review, in press
\rref Peebles, P. J. E., 1980, The Large Scale Structure of the Universe (Princeton: Princeton Univ. Press)
\rref Perlman, E. S., Padovani, P., Giommi, P., et al., 1998, AJ, 115, 1253
\rref Pogosyan, D., Bond, R., Kofman, L., Wadsley, J., 1998, preprint
\rref Ponman, A., et al., 1994, Nature, 369, 462
\rref Postman, M., 1998, in ``Evolution of Structure: from Recombination to Garching" (August 1998), 
A. Banday et al. eds., in press (astro-ph/9810088)
\rref Pye, J. P., Morley, J. E., Warwick, R. S., Pilkington, J., Micela, G., Sciortino, S., Favata, F., 1997, in 
``Cool Stars in     Clusters and Associations: Magnetic Activity and Age  Indicators" (G. Micela, R. 
Pallavicini, S. Sciortino eds.), Memorie SAIt 68, 1089
\rref Press, W. H., Schechter, O., 1974, ApJ, 187, 425
\rref Randich, S., 1997, in ``Cool Stars in Clusters and Associations: Magnetic Activity and Age Indicators" 
(G. Micela, R. Pallavicini, S. Sciortino eds.), Memorie SAIt 68, 971
\rref Rosati, P., Della Ceca, R., Norman, C., Giacconi, R., 1998, ApJ, 492, L21
\rref Rosati, P., 1998, in ``Widerref Field Surveys in Cosmology", Proceedings of XIVth IAP meeting, S. 
Colombi \& Y. Mellier eds., in press (astro-ph/9810054)
\rref Rosner, R., Golub, L., Vaiana, G. S., 1985, AR\&A, 23, 413
\rref Scharf, C. A,  Mushotzky, R. F., 1997, ApJ, 482, L13
\rref Schmidt, M., Hasinger, G., Gunn, J., et al., 1998, A\&A, 329, 495
\rref Schneider, D. P., Schmidt, M., Hasinger, G. et al., 1998, AJ, 115, 1230
\rref Sciortino, S. 1993, in ``Physics of Solar and Stellar Coronae", J. Linsky  and S. Serio (eds.),  Kluwer 
Academic Publishers,  221 
\rref Sciortino, S., Damiani, F., Favata, F., Micela, G., Pye, J.,  1998, Astron. Nachr., 319, 108 
\rref Sciortino, S., Favata, F., Micela, G., 1995, A\&A, 296, 370
\rref Seward, F.D., Mitchell, M., 1981, ApJ 243 736
\rref Shanks, T., Boyle, B. J., 1994, MNRAS, 271, 753
\rref Sidoli, L., et al., 1998, A\&A 336 L81
\rref Slezak, E., Durret, F., Guibert, J., Lobo, C., 1998, A\&AS, 129, 281
\rref Snowden, S. L., Petre R., 1994, ApJ, 436, L123
\rref Stocke, J. T., Morris, S. L., Gioia, I. M., et al., 1991, ApJS, 76, 813
\rref Tagliaferri, G., Cutispoto, G., Pallavicini, R., Randich, S., Pasquini, L., 1994, A\&A 285, 272
\rref Tananbaum, H., Tucker, W., Prestwich, A.,  Remillard, R., 1997, ApJ, 476, 83
\rref Vikhlinin, A., McNamara B. R., Forman, W., Jones, C.,  Quintana, H., \& Hornstrup, A., 1998a, ApJ, 
498, L21
\rref Vikhlinin, A., Forman, W., Jones, C., \& Murray, S., 1995, ApJ, 451, 553
\rref Vikhlinin, A., McNamara, B. R., Forman, W., Jones, C., Quintana, H., \& Hornstrup, A., 1998b, ApJ, 
502, 558.
\rref Vogeley, M. S., 1998, in ``Ringberg Workshop on Large-Scale Structure'' (astro-ph/9805160)
\rref Wang, Q. D., Connoly, A. J., Brunner, R. J., 1997, ApJ, 487, L16 
\rref Wang, Q., et al., 1991, ApJ 374 475 
\rref West, M. J., Jones, C., Forman, W., 1995, ApJ, 451, L5
\rref White, N., et al., 1995, in X-ray Binaries, eds. White, Parmar \& van den Heuvel
\rref Xia, X.-Y., et al., 1998, ApJ, 496, L9
\rref Zucca, E., et al., 1993,  ApJ, 407,  470

\newpage

{\bf Figure I.1}: Comparison of the survey area versus the limiting sensitivity. The baseline 
of the mission almost reaches the sensitivity of the AXAF deep survey over an area more 
than twenty larger. The confusion limit is well below the XMM confusion limit.

{\bf Figure II.1}: 30'x30' optical image from the DSS plates, with superimposed the X-ray contours 
from a pointed ROSAT observation.  There are 26 X-ray sources in this area.   20 are distant 
AGNs.  Sources 'A' and 'C'  are two clusters of galaxies at $z=0.36$ and $z=0.55$ respectively, 
while 'B' is a poorer group at $z=0.12$.

{\bf Figure II.2}:  The effective area covered on the sky by the three most representative X-ray deep cluster surveys, the 
EMSS (Henry et al. 1992), the RDCS (Rosati et al. 1998) and CfA (Vikhlinin et al.  1998a), compared to the areas of 
the WAXS Shallow and Deep surveys.   Note that the flux limits for the WAXS surveys are rather conservative 
(adapted from Rosati 1998).

{\bf Figure II.3}:  The limiting flux reachable as a function of the integration time for a typical cluster, at a given S/N 
ratio.

{\bf Figure II.4}:  Comparison of the local ($z<0.3$) and distant ($0.3<z<0.8$) cluster X-ray luminosity function.   Filled 
circles and the dot-dashed line represent the local XLF as measured by De Grandi et al. (1998), filled triangles and 
stars are, respectively, the RDCS (Rosati et al. 1998) and EMSS (Henry et al. 1992) results in the $0.3<z< 0.6$ 
range; lozenges are the RDCS estimates  for the $0.5<z<0.85$ bin.

{\bf Figure II.5}: The filled circles give the cluster X-ray luminosity function from the RDCS survey (Rosati et al. 1998), 
in three redshift bins at a median redshift (top to bottom) of 0.17, 0.31, 0.58.  The three XLF have been displayed 
with a vertical shift $\Delta \log(\phi)=2$ for clarity. The two panels refer to two different cosmological models: an 
Einstein-deSitter model, and a flat model with $\Omega_0=0.3$ and a cosmological constant. Note how the data from the more distant 
bin marginally overlap those at smaller redshifts (Borgani et al. 1998).

{\bf Figure II.6}: A preliminary estimate of the power spectrum of cluster density fluctuations from the REFLEX survey, 
currently the best available sample of X-ray selected clusters from the ROSAT All-Sky Survey.    Note how using 
clusters , the sampling of density fluctuations is optimal on large scales ($k<0.2$), i.e. specifically where more 
information is needed and where this becomes difficult when using galaxy redshift surveys. (Schuecker et al., in 
preparation; Boehringer et al. 1998).

{\bf Figura II.7}: Output of a hydrodynamic/n-body simulation of the formation of a cluster of galaxies and the 
surrounding filamentary structures, within a 32 Mpc size cube (Bode et al. 1998).  The pictures show different 
quantities projected from a slab of 8 Mpc deep at  the center of the cube. Clockwise from top-left, they show the 
distribution of (a) the dark matter, (b) the gas, (c) the X-ray emission, and (d) the corresponding temperature field.  
Note how well the cluster and surrounding groups in the X-ray image trace the overall mass distribution.

{\bf Figure II.8}:  All sky distribution in galactic coordinate of the 8593 Tycho stars detected as PSPC sources above the 
limiting flux of  $\sim 2 \times10^{-13}$ erg s$^{-1}$ cm$^{-2}$. The density enhancement at low galactic latitude is clearly visible as well as an 
evident enhancement that has been associated with a physical structure, the so-called Gould Belt. This has an 
ellipsoidal shape with a semi-major axis of about 500 pc and a semi-minor axis of about 340 pc, and with the Sun 
located inside it about 150 to 250 pc off center (from Guillout et al. 1998a).

{\bf Figure II.9}: A simulation of the projected all-sky angular distribution (in galactic coordinates) of the X-ray emitting 
stars down to $m_V = 15$ and to $f_X  = 5 \times10^{-14}$ erg s$^{-1}$ cm$^{-2}$ according to the model of the spatial distribution of young 
stars recently proposed by Guillout et al.(1998b). Note that both the galactic disk population and the Gould belt 
populations will be adeguatelly sampled by the proposed low-latitude surveys (Simulation is courtesy of P. Guillout).

{\bf Figure II.10}: Detection by the wavelet algorithm on a simulation of an ultra deep WAXS field. The size of the circles is 
related to the physical dimension of the detected sources.

{\bf Figure II.11}: Comparison between the input $\log N - \log S$ and that derived from the analysis.

{\bf Figure III.1}: View of the WFXT parts.

{\bf Figure III.2}: WFXT Mirror Module conceptual design.

{\bf Figure III.3}: Image spot of the outermost WFXT mirror shell tested at the PANTER facility. The four images (from 
left to right and top to bottom) refer to off-axis positions of 0', 10', 20' and 30' off-axis angles. Images were taken at 
energy of 1.5 keV and with a CCD detector.
 
{\bf Figure III.4}: Effective area (mirror + CCD + filter) for the WFXT telescope. Different curves refer to the different 
material used to coat the mirror surfaces.

\end{document}